\documentclass{pja00}
\usepackage{graphicx}
\usepackage{color}
\input ulem.sty

\begin{document}

\runninghead{Subaru Studies of the Cosmic Dawn}

\title{Subaru Studies of the Cosmic Dawn} 

\author{Masanori Iye}
\affiliation{Thirty Meter Telescope Project, National Astronomical 
Observatory, Mitaka, Tokyo 181-8588}

\Subject*{}

\maketitle

\begin{abstract}

An overview on the current status of the census of the early universe population is given.  Observational surveys of high redshift objects provide direct opportunities to study the early epoch of the Universe.   The target population included are Lyman Alpha Emitters (LAE), Lyman Break Galaxies (LBG), gravitationally lensed galaxies, quasars and gamma-ray bursts (GRB).  The basic properties of these objects and the methods used to study them are reviewed.  The present paper highlights the fact that the Subaru Telescope group made significant contributions in this field of science to elucidate the epoch of the cosmic dawn and to improve the understanding of how and when infant galaxies evolve into mature ones.  
\end{abstract}

\section{INTRODUCTION}
Since the discovery of the expansion of the Universe by Edwin Hubble in 1929, astronomers attempted to look at distant galaxies with increased power of new generation telescopes. By studying distant galaxies, one can look back the early history of the Universe. Partridge and Peebles (1967)\cite{1}, in their classical paper, predicted the properties of primordial galaxies and pointed out that these galaxies with redshifted Ly$\alpha$ emission would be the targets observational astronomers should look for.  As was reviewed in Pritchet(1994)\cite{2}, many attempts followed using 4m class telescopes for the next three decades without leading to a great success. 

National Astronomical Observatory of Japan built the 8.2m Subaru Telescope atop Mauna Kea during 90's, which became in full operation by 2002\cite{3}.  Astronomers in the world started to find out extremely distant objects using 10m Keck telescope, 8.2m Subaru Telescope, 8.1m VLT, and 8.0m Gemini Telescopes.  The observational techniques used are 1) narrow band imaging surveys for Ly$\alpha$ emitting galaxies\cite{4}-\cite{29}, 2) multi-band photometric surveys for Lyman break galaxies\cite{30}-\cite{38},  3) searches for amplified images of gravitationally lensed galaxies\cite{39}-\cite{44}, 4) photometric surveys for high redshift quasars\cite{45}-\cite{47}, and 5) studies of sporadic gamma ray bursts\cite{48}-\cite{51} in high redshift galaxies. As of today, the highest spectroscopically confirmed redshift records are $z=7.213$ for LAEs\cite{29}, $z=7.085$ for quasars\cite{47}, and $z\sim8.2$ for GRBs\cite{51}.   There are additional candidates of Lyman break galaxies whose photometric redshift estimate indicate $z>$8\cite{35}-\cite{38}. 

The current picture of the big bang Universe indicates that the expanding universe cooled rapidly to form neutral hydrogen from protons and electrons at 380,000 years after the big bang. This is the epoch when the photons are decoupled from the matter and constitutes the last scattering surface.  Hence, this is also the epoch of the Universe beyond which observation by electromagnetic radiation cannot penetrate. Figure 1 shows the power spectrum of the temperature fluctuation at this epoch derived from 7 years of WMAP observation of the cosmic microwave background\cite{52,53}. Measured data showing the baryon acoustic oscillation features and the onset of Silk damping tail are very well reproduced by a $\Lambda$CDM model, constraining the cosmological parameters, cosmic age $t_{0}=13.75\pm0.13$ Gyr, the Hubble constant$H_{0}=71.0\pm2.5$ km/s/Mpc, baryon density $\Omega_{b}=0.0449\pm0.0028$, cold dark matter density $\Omega_{c}=0.220\pm0.026$, and vacuum energy density $\Omega_{\Lambda}=0.734\pm0.029$. The density fluctuation of the dark matter and the matter grew by gravitational instability and it is conceived that the first generation of stars were born at around 200 million years after the big bang. 

The initial set of formed stars contained wide range of mass spectrum. As all the elements heavier than boron were produced through the nuclear burning of hydrogen inside stars and were not existent in the early universe where only hydrogen and helium were formed during the first 5 minutes of the big bang, astronomers use a brutal term, "metal" for all the elements heavier than carbon.  The absence of "metal" elements in the primordial gas helped to form massive stars. Due to the strong UV radiation from those newly formed massive hot stars, the surrounding intergalactic matter was gradually re-ionized.  It is like a "Global Warming" of the Universe.  When and how these re-ionization process took place was not observationally clarified yet but polarization measurement of WMAP7 \cite{53} suggests that it took place at z$\sim10.5\pm1.2$, if the re-ionization was an instantaneous event. It is more likely that the cosmic re-ionization could have taken place in an extended period sometime during 6 $<$ z $<$17. 

Detailed observations deep into the era beyond $z=7$ is, therefore, crucial.  Some of the recent number counts of galaxies at 5.7 $<$ z $<$ 7 indicate significant decrease in the number density of Ly$\alpha$ emitting galaxies, especially at its bright end of the luminosity function\cite{13}-\cite{19}, which could either be due to the premature formation of those bright galaxies through merging processes or due to the increasing fraction of neutral hydrogen that blocks the visibility of Ly$\alpha$ emitting galaxies at high redshift. 

In the subsequent sections 2-6, I will review the target population of galaxies in the early Universe and the technique astronomers are employing to find those objects together with some recent results.   The summary will be given in the last section.

\begin{figure}
   \begin{center}
      \includegraphics[width=7.5cm]{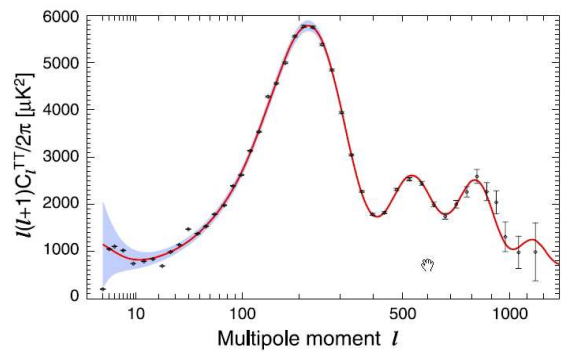}
   \end{center}
   \caption{Temperature fluctuation power spectrum from 7 years observation by WMAP (Larson et al.(2011)\cite{53}).  The ordinate shows the auto correlation of the expansion coefficients of the cosmic microwave background temperature distribution in spherical harmonics $Y_{l}^{m}$.  Measured data points are very well reproduced by a $\Lambda$CDM model denoted by a solid curve, that constrains many of the cosmological parameters.  The gray band represents the uncertainty due to the cosmic variance.}
   \end{figure} 
\vspace{5mm}

\section{NARROW BAND SURVEY FOR LYMAN ALPHA EMITTERS}

Ly$\alpha$ emitters, often abbreviated as LAEs, are thought to be young galaxies in star-forming phase lasting about 10 million years with star formation rate from 1 to 10 solar mass per year.   Hot massive stars produce strong UV radiation field and ionize the interstellar gas. The ionized hydrogen recombines and cools by emitting finally a Lyman $\alpha$\rm photon to settle down to the lowest ground level.  The amount of stars produced in LAEs is yet of the order of $10^{8}$ solar mass and the continuum radiation from stars is not necessarily conspicuous.  The spectra of LAEs are therefore characterized by strong Lyman-alpha emission line as shown in Figure 2.  At high redshift universe beyond $z\sim5$, due to the absorption of Ly-$\alpha$ photons by the residual neutral hydrogen in the line of sight to us, the blue part of the emission line profile is truncated and LAEs show a characteristic asymmetric profile. The identification of the observed emission line to Ly-$\alpha$ 121.5nm line at high redshift, rather than other typical emission lines, e.g., C{\sc iv} 154.9nm, Mg{\sc ii} 279.8nm, [O{\sc ii}] 372.7nm, and [O{\sc iii}] 500.7nm, at lower redshifts, can be properly made by covering a wide enough spectral range to check the presence of other possible lines.  However, even for a single line, the presence of this asymmetric emission line profile provides fair confidence to the line identification.

   \begin{figure}
   \begin{center}
      \includegraphics[width=7.5cm]{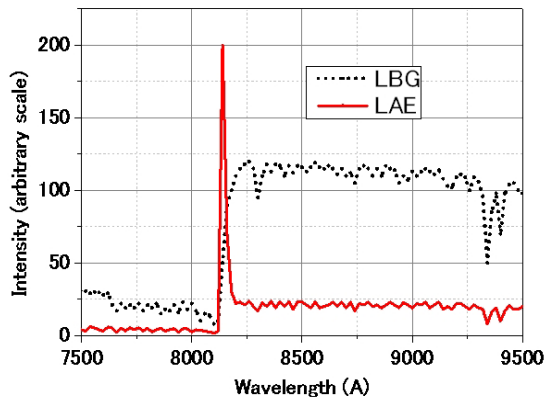}
   \end{center}
   \caption{Typical spectra of a Lyman-Alpha Emitter (solid line) and a Lyman Break Galaxy.(broken line) at an assumed redshift 5.7, showing the Ly$\alpha$ emission line at 815nm and stellar continuum long ward of the Lyman $\alpha$ emission line.}
   \end{figure} 

Although LAEs are characterized by their redshifted Ly$\alpha$ emission line, they are so faint and the Earth's foreground night sky emission, that consists of emission line bands of OH radicals in the upper atmosphere and the thermal blackbody emission, hinder their detection in the near infrared spectral region, $\lambda>680$nm, where the Ly$\alpha$ emission is shifted for galaxies with redshift $z>4.6$.   Figure 3 shows the strong bands of telluric OH emission lines in the wavelength region below 1.05 micron, where Si-CCDs are sensitive.  Astronomers use the dark gaps between these OH bands to probe deep space.  

Subaru observational astronomers formed a consortium to develop a series of narrow band filters for the wide field camera, Suprime\textcolor
{red}{-}Cam \cite{54}, whose transmittance bands are matched to one of these OH band gaps to detect LAEs whose redshifted Ly$\alpha$ emission enters in such dark sky windows.  LAEs at appropriate redshift range are expected to show up brighter in the narrow band image than in other broad band images.  The narrow band (NB) survey is therefore trying to slice the universe in a narrow range of redshift. For instance, the narrow band filter NB816 that has the central wavelength at 816nm is suitable for isolating LAEs at $z=5.7$, NB921 for $z=6.6$, etc. The most distant LAE at $z=6.964$ was also discovered using another filter NB973\cite{15} . The replacement of Suprime-Cam CCDs to new red sensitive ones with thicker depletion layer in 2009 provided enhanced sensitivity at around 1 micron and the reddest filter NB1006 for redshift 7.3\cite{18} was also added to the Suprime-Cam filter suite.

   \begin{figure}
   \begin{center}
      \includegraphics[width=7.5cm]{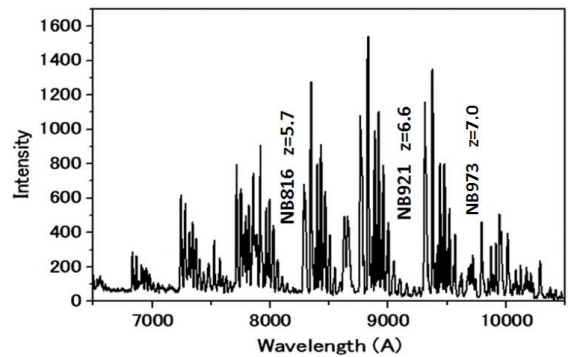}
   \end{center}
   \caption{OH night sky emission bands showing a few gaps, which astronomers use as dark windows to look deeply into the Universe.   Narrow band filters whose transmission are matched to these dark windows are used to sample LAEs at $z$=5.7 (NB816), $z$=6.6 (NB921) and $z$=7.0 (NB973). }
   \end{figure} 

The Subaru Deep Field survey\cite{55} and the Subaru XMM-Newton Deep Field survey\cite{56} were designed to enable systematic studies of galaxy population in the early universe by using the Suprime-Cam imaging data obtained in broad bands, $B, V, R, i'$ and $z'$ to allow photometric redshift estimate of each object from their observed spectral energy distribution fitted to stellar population synthesis models for galaxies.  This unique data set sampling the large space volume in the early universe provided a reference frame to study the luminosity function, morphological and clustering properties, time evolution of galaxies in the early universe.   Narrow band imaging data using NB711, NB816,  and NB921 added  precious samples for studying LAEs at $z\sim4.8, 5.7$ and 6.6, respectively. These data are also used to study star forming galaxies and active galactic nuclei at lower redshifts.

   \begin{figure}
   \begin{center}
      \includegraphics[width=7.5cm]{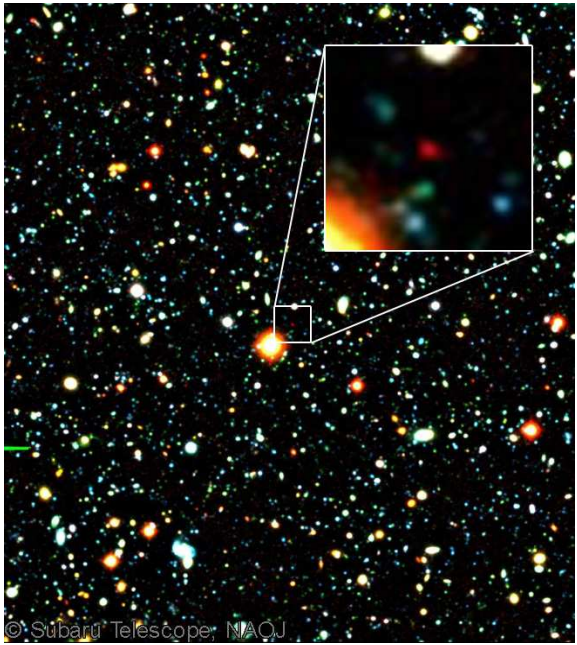}
   \end{center}
   \caption{The most distant galaxy IOK-1 is shown as a red blob in the inlet panel covering $8\times8$ arcsec in the sky.  The entire field of view covers $254\times284$ arcsec.  North is up and East to the left.}
   \end{figure}

   \begin{figure}
   \begin{center}
      \includegraphics[width=7.5cm]{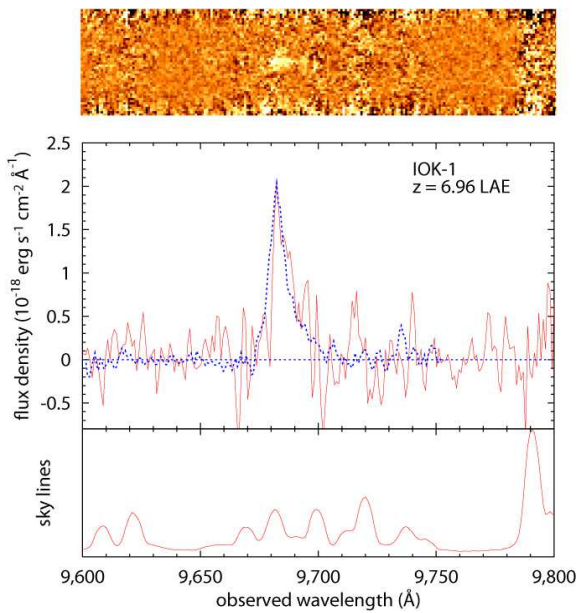}
   \end{center}
   \caption{The red line in the center panel shows the observed spectrum of IOK-1, showing the characteristic Ly$\alpha$ emission line with an asymmetric profile at 968nm indicating its redshift 6.96 (Iye et al.(2006)\cite{15}). The blue line superposed shows the averaged template profile of Ly$\alpha$ emission lines for LAEs at lower redshift shifted to $z=6.96$.  The red line in the lower panel shows the foreground OH night sky emission lines in the same spectrum.}
   \end{figure} 

The red blob in the inlet panel of Figure 4 shows the most distant LAE, IOK-1, at the time of its discovery by Iye et al(2006)\cite{15} and remained at the top of the list until the end of 2010 when two LAEs were found beyond $z=7$\cite{25}.   IOK-1 was discovered among the 41,533 objects above 5$\sigma$ threshold detected in the Subaru Deep Field through the narrow band filter NB973 for a total exposure time of 15 hours with  Suprime-Cam. All the objects were cross identified in images taken in other filters and five objects, IOK-1 to IOK-5, which are visible only in this narrow band filter, were identified as photometric candidates for $z\sim7$ LAEs. 

Color excess in NB973 is a necessary, but not sufficient, condition for $z\sim7$ LAEs.  There are several types of possible contaminants in these 5$\sigma$ photometric candidates. First, since the narrow band imaging observation was made 1-2 year after other broad band observations, some of the candidates may well be variable objects like AGNs or galaxies where supernovae added extra light when narrow band observation was made.  Possibility for emission line objects at a lower redshift is a common concern. To our surprise, simple Gaussian statistics cautions us that there might be one or two 5$\sigma$ "noises" as well, since there are millions of independent 2 arcsec  apertures one can sample in the Suprime-Cam field.  Therefore spectroscopic follow up observation is essential to establish a firm confirmation.  Eventually, it was revealed that only one object, the brightest IOK-1, is a real LAE at redshift 6.96, with the characteristic asymmetric line profile as shown in Figure 5.


\begin{table}
\caption{Top 21 distant galaxies (Aug. 8, 2011)}
    \begin{tabular}{lllll}
\hline
$\#$&ID&$z$&Paper&Date\\
\hline\hline
1&GN-108036&7.213&Ono+&2011.7\\
2&BDF-3299&7.109&Vanzella+&2011.4\\
3&A1703zD6&7.045&Schenker+&2011.7\\
4&BDF-521&7.008&Vanzella+&2011.4\\
5&G2-1408& 6.972&Pentricci+&2011.7\\
6&IOK-1&6.964&Iye+&2006.9\\
7&HUDF09-1596&6.905&Schenker+&2011.7\\
8&SDF46975&6.844&Ouchi+&2011.7\\
9&NTTDF6345&6.701&Pentricci+&2011.7\\
10&NTTDF474&6.623&Pentricci+&2011.7\\
11&SXDN71598&6.621&Ouchi+&2010.11\\
12&SXDW30717&6.617&Ouchi+&2010.11\\
13&SDF1004&6.597&Taniguchi+&2005.2\\
14&SDF1018&6.596&Kashikawa+&2006.4\\
15&SXD Himiko&6.595&Ouchi+&2008.7\\
16&SXDW31790&6.590&Ouchi+&2010.11\\
17&SXDC106098&6.589&Ouchi+&2010.11\\
17&SDF1030&6.589&Kashikawa+&2006.4\\
19&SDF91163&6.587&Kashikawa+&2009.2\\
19&SDF91988&6.587&Kashikawa+&2009.2\\
19&SDF71101&6.587&Kashikawa+&2009.2\\

\hline
 \end{tabular}\\
   
\end{table}

Table 1 shows the top 21 list of high redshift galaxies with spectroscopic redshift confirmation to the best of the authors knowledge at the time of writing this review.  Although, a discovery of redshift 8.6 LAE was reported in 2010\cite{25}, the published data do not appear as a robust one and I haven't included this object in this list. Readers may notice that 14 out of 21 were discovered by Subaru/Suprime-Cam surveys.  This is because Subaru/Suprime-Cam enables observation of large survey volume with significant depth.   

   \begin{figure}
   \begin{center}
      \includegraphics[width=7.5cm]{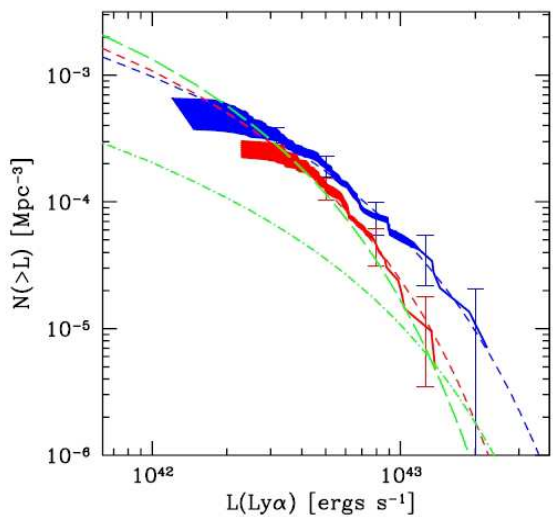}
   \end{center}
   \caption{Ly$\alpha$ luminosity functions of LAEs at $z$=5.7 (blue) and $z$=6.6 (red).  Note that the significant decrease in Lyman-alpha luminosity function at its bright end (Kashikawa et al., 2011\cite{23}). }
   \end{figure}

Subaru Deep Field surveys yielded several dozens of LAE candidates both at redshift 5.7 and 6.6, of which more than half of them are already confirmed spectroscopically to be LAEs. With this fair sample, one can derive the luminosity function of LAEs.  It was confirmed that the UV continuum luminosity functions of LAEs at redshift 5.7 and 6.6 are practically identical.  On the other hand, figure 6 shows that the Ly$\alpha$ luminosity function derived at redshift 6.6 has a significant deficiency with respect to that at redshift 5.7 especially at their brighter population\cite{23}.  

This can be explained if the neutral hydrogen fraction of the intergalactic matter is increasing from redshift 5.7 to 6.6, as the neutral hydrogen selectively absorbs and scatters the Ly$\alpha$ photons but not affects UV continuum photons. Such a change in the fraction of neutral hydrogen is possible by the ionizing photons produced by the first generation of massive hot stars.  Actually, the observed Ly-$\alpha$ luminosity functions, the UV luminosity functions, and the distribution of equivalent width of the Ly-$\alpha$ emission are compatible with a star formation history where an initial Pop III massive star formation was followed later by a Pop II star formation\cite{57}.

Recent follow up HST imaging of IOK-1 in the F130N, F125W, and F160W filters to probe the redshifted He{\sc ii} $\lambda$1640 emission placed an upper limit for the amount of metal free Pop III stars\cite{25}.
The scarcity of massive galaxies at higher redshift could be ascribed, in principle, to premature evolutionary process of galaxies built from tiny proto galaxies.  However, in this case, both the LAE luminosity function and the UV continuum luminosity function should show a coherent difference.  Cosmic variance, though not negligible, could not be the main cause of this striking difference.  

The significant change in the fraction of neutral hydrogen during $6 < z <7.2$ is also supported by recent additional works\cite{28,29,47}.
We, therefore, conclude that observations of the galaxy population in the early universe started to reveal the epoch when the cosmic re-ionization was not completed.   This last phase of cosmic re-ionization is often referred as "cosmic dawn".  

The upper panel of Figure 7 shows the angular correlation function for LAEs at $z=6.6$ as a function of angular separation derived for the first time in the Subaru -XMM deep Field\cite{20}.  Filled squares are observed values, which is fitted by a solid line.  The existence of positive correlation at shorter scale implies the presence of clustering.  Dotted line is that for underlying cold dark matter predicted by a theoretical model. 

   \begin{figure}
   \begin{center}
      \includegraphics[width=7.5cm]{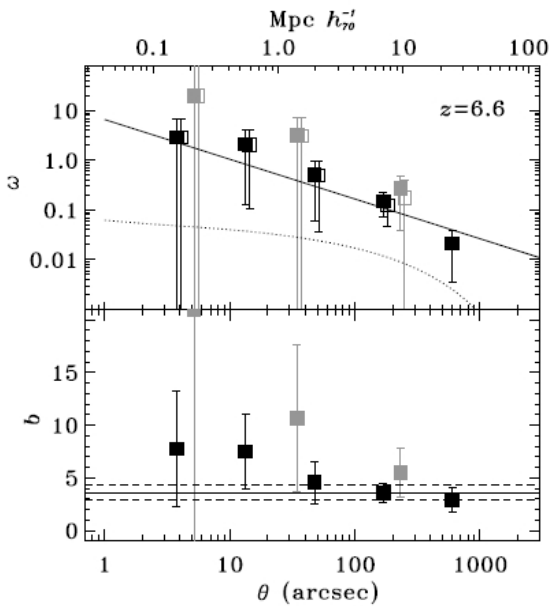}
   \end{center}
   \caption{Angular correlation function $\omega$ and bias $b$ as a function of angular distance $\theta$ derived for 207 LAEs discovered in the Subaru-XMM Deep Field.  Black and gray squares represent measured values for all and bright subsamples ,respectively. Solid line in the top panel shows best fit power-law, while the dotted line denotes the angular correlation function of the underlying dark matter predicted by a CDM model. The evidence of clustering is shown for the first time at $z=6.6$ (Ouchi et al. 2011\cite{24}.)}
   \end{figure} 


   \begin{figure}
   \begin{center}
      \includegraphics[width=7.5cm]{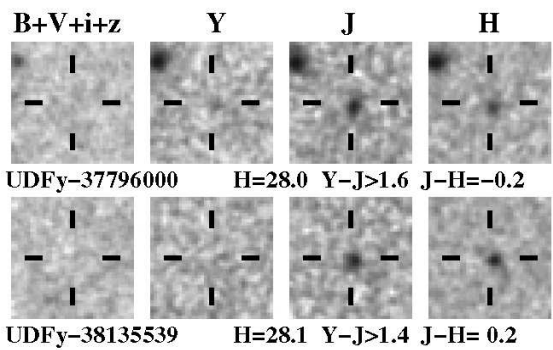}
   \end{center}
   \caption{Lyman break galaxy candidates at $z\sim8.6$ discovered as Y-dropouts from the Hubble Ultra Deep Field (Bouwens et al. 2011\cite{38}.)}
   \end{figure}

\vspace{5mm}

\section{TWO COLOR DIAGNOSIS FOR LYMAN BREAK GALAXIES}

Another population of galaxies searched for in the early Universe is called Lyman Break Galaxies, abbreviated as LBGs. LBGs are thought to be fairly massive galaxies with evolved stellar population. Stellar continuum is much stronger than LAEs.  Ly$\alpha$ emission is less conspicuous as compared with LAEs.  The spectra of these galaxies show characteristic discontinuity at the blue side of Ly$\alpha$ line caused by the intrinsic stellar atmospheric absorption and by the intergalactic neutral hydrogen absorption (cf. Figure 2).  These galaxies, therefore, are visible at bands red ward of Ly$\alpha$ line but are not visible at bands blue ward of the Ly$\alpha$ line. One can select out LBG candidates at $z$=6 by looking for objects that is not visible in the i-band and other bands at shorter wavelengths but is visible in bands at longer wavelengths, which is called "i-band dropouts".  Similarly, one can look for $z$=7 LBGs as $z$'-band dropout, and $z$=9 LBGs as J-band dropouts.

Here again, one have to be careful for possible contaminants. Galactic T-dwarfs, a class of low-mass, low-temperature, low-luminosity stars\cite{58}, dwell in the similar region in the two color diagram.  One may be able to reject T-dwarfs by their point source images if the image quality is superb.   Variable objects and 5$\sigma$ noises are the common problems for this survey as well.

Hubble ACS and NICMOS imaging at Hubble Ultra Deep Field and GOODS field was used to identify faint $z$-dropouts at around $z$=7.3 and about 8 candidates were identified, but similar attempt for J dropout didn't yield a candidate\cite{33}. Another group reported finding of 10 $z$-dropouts and 2 J-dropouts\cite{33}.  Unfortunately, many of these objects do not show strong Ly$\alpha$ emission and spectroscopic confirmation of their genuine redshift remains difficult with currently available 8-10m class telescopes.

Figure 8 shows two examples of LBG candidates at $z\sim8.6$ discovered as Y-dropouts from the Hubble Ultra Deep Field survey\cite{36}.  They are not visible in filter bands at $B$, $V$, $R$, $i$, $z$, and $Y$, but detected in infrared $J$ and $H$ bands.  Photometric redshift estimates based on comparison of the derived spectral energy distribution with those models of galaxies place them possibly at $z\sim8.6$.

Hubble Ultra Deep Field imaging survey with ACS probed much deeper than ground based observations, but has a much smaller survey volume.  The wide field surveys to pick up scarce bright population and narrow field deep surveys to study fainter populations, are complementary to each other.

\vspace{5mm}
\section{SURVEY FOR STRONGLY LENSED GALAXIES}

Another unique survey project is to take advantage of the gravitational lensing effect of a massive cluster of galaxies, which will magnify and brighten the background faint galaxies. In this sense, clusters of galaxies are the largest cosmic telescopes in the Universe with a diameter about
1Mpc.  They are in situ and free of charge to use. However, you cannot point them to your favorite targets. Wavefront aberrations are bizarre. Although the images produced by cluster lensing are peculiarly deformed and enlarged, the largest advantage is the fact some of the lensed images are brightened considerably and when multiple lensed images are available they can be used to check for the consistency of their reconstructed source image.

  \begin{figure}
   \begin{center}
      \includegraphics[width=7.5cm]{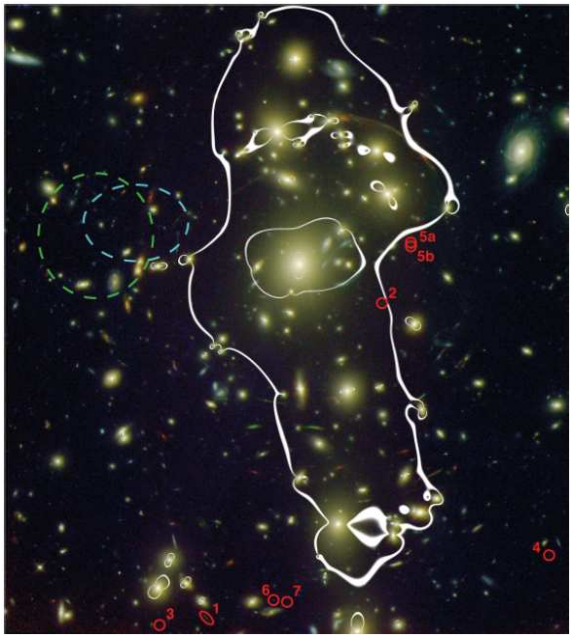}
   \end{center}
   \caption{Lyman break galaxy candidates at $z\sim7$ marked in red circles and ellipses as discovered behind the gravitational lensing cluster A1703 at $z=0.28$ (Edited from Bradley et al. 2010\cite{44}.)}
   \end{figure}

Appropriate modeling of the gravitational field of the cluster enables the prediction of the location of critical lines for assumed source redshift slice where the magnification becomes infinity. Observers can look for lensed object along these critical lines and there are in fact several candidate galaxies found in this way\cite{39}-\cite{44}.  For instance, a survey for strongly lensed LAEs in 9 clusters yielded six candidates\cite{40}.   If any of these candidates are real, the number density of faint population of galaxies is much larger than previously considered and may well explain the necessary amount of re-ionizing source.

Figure 9 shows several $z$-dropout objects that could be Lyman Break Galaxies at redshift $z\sim7$ behind the cluster Abel 1703 at $z=0.28$\cite{44}.   Photometric results indicate better match to galaxies at $z\sim7$, however, here again the possibility of galaxies at $z\sim1.7$ is hard to rule out just from imaging.

\vspace{5mm}
\section{QUASARS}

The next category of objects of interest in the distant, early universe are point sources, first, quasars. The photometric survey technique used to isolate high redshift quasar candidates is similar to that used for LBGs. Objects that match the expected spectral energy distribution of high redshift quasars are surveyed in the two color diagram or even a multi-dimension color manifold.   Sloan Digital Sky Survey with its enormous data base is a nice test bed to apply this approach. Many quasars beyond redshift 6 were found in this way\cite{45}-\cite{47}. Gunn-Peterson tests\cite{59,60} of quasars up to redshift 6 indicate strongly that the cosmic re-ionization ended by redshift 6. The most distant quasar found to date is ULAS J1120+0641 at 7.08\cite{42}, which also supports the increase of neutral hydrogen fraction from $z=6$ to $z=7$.

\vspace{5mm}
 
\section{GAMMA RAY BURSTS}   

The advent of the real time alert system of gamma ray burst increased the chance for optical and infrared astronomers to make prompt observations of these rapidly declining bursts\cite{48}-\cite{51}.  The most distant GRB observed to date is GRB090423 at $z\sim8.2$\cite{51}.  GRBs at high redshift can be useful tools to probe the cosmic re-ionization through its Lyman-alpha damping wing\cite{51}.

Timely spectroscopy of GRB050904\cite{48,49} using FOCAS\cite{61} was very successful in deriving an estimate on the intergalactic neutral hydrogen from its damped Ly$\alpha$ wing absorption profile as well as giving a concrete evaluation of the chemical abundance in its host galaxy at $z\sim6.3$.

GRB has a much simpler featureless continuum than the quasar spectra which has broad emission lines superposed on the non-thermal continuum. GRBs are, in a way, better probes to study the re-ionization history.  Both quasars and GRBs are point sources, the advent of laser guide star adaptive optics makes the observation of fainter objects feasible and we expect many such observations will be conducted soon, if the observatories pay efforts for timely follow-up spectroscopy of long burst GRBs.  GRBs may provide a new way to study even higher-redshift galaxies and first generation of stars. 

\vspace{5mm}

\section{CONCLUSION}

The 8.2m Subaru Telescope\cite{3}, operated by the National Astronomical Observatory of Japan, at the summit of Mauna Kea has made a systematic census of early galaxy populations, especially the Ly$\alpha$ emitters for which a precise redshift measurement is feasible, in the Subaru Deep Field\cite{55} and Subaru-XMM Deep Field\cite{56}, using a series of narrow band filters mounted on the wide field camera, Suprime-Cam\cite{54}, a unique instrument among those for 8-10m class telescopes.   Owing to its wide field coverage of Suprime-Cam and a dedicated consortium to make a systematic observation feasible, the Subaru Suprime-Cam data base was extremely successful in finding many of distant populations of galaxies beyond redshift 6. The detectability of Ly$\alpha$ emission has been used as an important criterion to diagnose the epoch of the cosmic re-ionization\cite{57}.

Fig.10 shows the increase of the fraction of neutral hydrogen as measured from Gunn-Peterson tests\cite{59} of quasars up to redshift 6.42 on the left hand. Our results from redshift 6.6 and 7.0 LAE is shown in red and an upper limit from redshift 6.3 GRB is shown in blue triangle. Planck satellite may give more clue in 5 years time.  Surveys for galaxies beyond redshift 7 up to 11 is, therefore, extremely important to elucidate what happened actually in this period and for that we need NIR deep surveys.

   \begin{figure}
   \begin{center}
      \includegraphics[width=7.5cm]{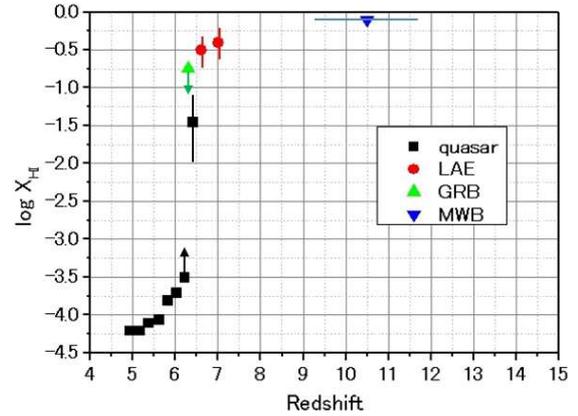}
   \end{center}
   \caption{Neutral hydrogen fraction of intergalactic matter as derived from Gunn-Peterson tests of $z>$5 quasars (black squares), damped Lyman-alpha wing profile (blue triangle), and Ly$\alpha$ luminosity function (red circles).  Also plotted is the WMAP 7 year result, which predicts $z\sim10.5\pm1.2$ for instantaneous re-ionization\cite{53}.  Note, however, that WMAP cannot constrain when re-ionization started and how long it took to complete. }
   \end{figure}

   \begin{figure}
   \begin{center}
      \includegraphics[width=7.5cm]{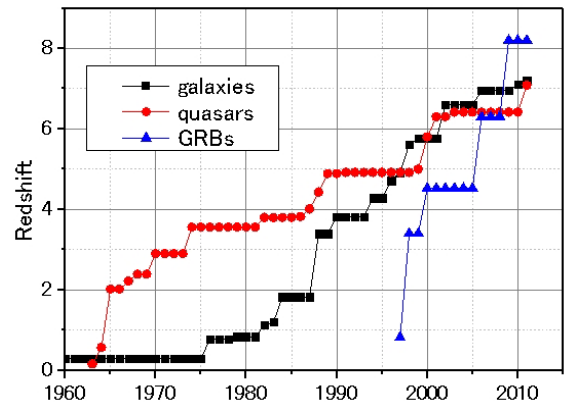}
   \end{center}
   \caption{Growth history of largest redshift objects. Note that GRBs are catching up quickly.
}
   \end{figure}    

Fig. 11 shows the annual growth of the records of the highest redshift objects.  The Subaru discovery of $z$=6.964 galaxy, IOK-1, was announced on Sep.14, 2006, and remained at the top until October 2010.  As shown in Table 1, several galaxies at $7.0<z<7.2$ were discovered during the last 10 months.  The highest redshifts of quasars and GRBs are also exceeding $z=7$.  Some LBGs are also found whose photometric redshift estimate indicate $z\sim 8-10$.  To extend the survey volume at high redshift Universe, Hyper Suprime-Cam\cite{62} having 7 times wider field of view than Suprime-Cam and the Large Synoptic Survey Telescope (LSST)\cite{63} will play important roles.   We are about to start seeing the cosmic dawn but to make a great leap in this science, we do need next generation of space telescope, James Web Space Telescope (JWST)\cite{64}, to probe population $z>10$.  The next generation ground based extremely large telescopes, Thirty Meter Telescope (TMT)\cite{65}, European Extremely Large Telescope (E-ELT)\cite{66}, and Giant Magellan Telescope (GMT)\cite{67} will elucidate the concrete characteristics of the infant Universe.

\section{ACKNOWLEDGMENT}

The author expresses his sincere respect and acknowledgment to all the individuals who contributed to the planning, construction, and operation of Subaru Telescope and those who made great use of this facility to advance the studies of the universe as reviewed in this article. He is also grateful to anonymous referees who reviewed the draft with great care and made useful suggestions.

\end{document}